**Session VI. TOWARDS THE EXPLORATION OF NEW NICHES AS RESERVOIRS OF MICROBIAL DIVERSITY**

# Text-mining and ontologies: new approaches to knowledge discovery of microbial diversity

NEDELLEC Claire (1) [*], BOSSY Robert, CHAIX Estelle, DELEGER Louise

(1) MaIAGE, INRA, Université Paris-Saclay, 78350 Jouy-en-Josas, France
[*] **Corresponding Author**:   claire.nedellec@inra.fr

**Introduction**
Microbiology research has access to a very large amount of public information on the habitats of microorganisms. Many areas of microbiology research uses this information, primarily in biodiversity studies. However the habitat information is expressed in unstructured natural language form, which hinders its exploitation at large-scale. It is very common for similar habitats to be described by different terms, which makes them hard to compare automatically, e.g. intestine and gut. The use of a common reference to standardize these habitat descriptions as claimed by (Ivana et al., 2010) is a necessity. We propose the ontology called OntoBiotope that we have been developing since 2010. The OntoBiotope ontology is in a formal machine-readable representation that enables indexing of information as well as conceptualization and reasoning.

**Background**
Microorganism habitats are described in a wide variety of sources and for multiple reasons. All areas of fundamental and applied microbiology produce habitat descriptions, primarily in the form of articles. Moreover, biological resource databases have always included a more or less structured and detailed fields for isolation site information (*e.g.* BacDive, *the Bacterial Diversity Metadatabase* of DSMZ[1] with 24150 "isolated from" entries or ATCC, the *American Type Culture Collection* with 18000 isolation information[2]). They are collected by aggregator sites, (*e.g. Global Biodiversity Information Facility* (GBIF)[3], *StrainInfo*[4]).
New technologies for the identification of microorganisms by DNA-based technologies, such as high throughput sequencing methods generate a huge number of sequences of microorganisms for a wide variety of environments that are stored with their isolation site in large databases such as GenBank or the *Sequence Read Archive* (SRA). This results in exponential growth in the volume of publications and database entries. 7 million microbial habitats are mentioned in PubMed references according to (Deléger et al., 2016). More than 25000 habitats of bacteria and archae are stored in the JGI *Genome OnLine Database* (GOLD)[5].
At the same time, the abundance of descriptions encourages the emergence of cross-cutting issues such as questions related to the origin and contamination routes or the adaptation of microorganisms to different environments in relation to evolutionary and genetic issues. The processing of such a volume of data requires the use of automatic methods. As the field of bioinformatics provides answers to the processing of structured data, sequences, structures, *etc.*, the descriptions of environments remain largely underutilized due to a lack of solutions. The reason is twofold. The large-scale analysis of descriptions of the living environments of microorganisms requires (1) a reference classification to which to link descriptions and (2) an automatic means of associating descriptions of habitats and categories. This second point falls under text-mining, which involves the use of automatic language processing and learning methods to finely extract information, categorize it with the reference classification and finally link it to the organism when needed (Nédellec et al. 2009). Recent advances in text-mining have made it possible for methods to achieve high level performances that make them usable nowadays for microbiology biodiversity study, as measured for example at the BioNLP Shared Task competition for microbial habitat categorization (Bossy et al., 2013).

---
[1] https://bacdive.dsmz.de
[2] https://www.lgcstandards-atcc.org/Search_Results.aspx?dsNav=Ro:0,N:1000552&adv=1&redir=1
[3] https://www.gbif.org
[4] http://www.straininfo.net
[5] https://gold.jgi.doe.gov/metadatasearch

This paper addresses the first point, the availability of a classification of reference habitats. To be usable, the classification must meet a number of criteria. On the one hand, it must be rich enough to reflect the great diversity of microbial habitats and to distinguish habitats with different physico-chemical properties. On the other hand, it should not be too broad, which would be detrimental to its maintenance and manual use. Its structure must both reflect the areas of microbial biodiversity studies to facilitate its appropriation by microbiologists and group similar environments in order to facilitate their processing. Its organization must be hierarchical to allow its use at different levels of precision. A given category could belong to more than one category in order to allow description from several points of view, but this case must remain marginal for the structure to remain readable and compact.

Classifications of microorganism habitats are few and do not fully meet these criteria. For example, the ATCC classification is a list of 37 environmental habitat entries (Table 1 of (Floyd et al., 2005)), insufficient for microbial biodiversity study because of its small size and flat structure. GOLD uses a richer but flat controlled vocabulary (Reddy et al., 2014) to index the isolation information of the sample. EnvO (*Environment Ontology project*)[6] is a rich classification of 7000 classes supported by the *Genomics Standards Consortium* (GSC). It aims to support standard manual annotations of environments and biological samples (Buttigieg et al., 2013). However, it suffers from some limitations for microorganism biotope descriptions. It fails to account for main trends in microorganism studies, such as technological use for food transformation or bioremediation, and their pathogenic or symbiotic properties. The principle of EnvO design relies on the reuse of well-known schemes with the consequence that they may not be fully appropriate for habitat categorization. Thus, EnvO soil branch from the Agriculture Organization soil classification (Buttigieg et al., 2013) includes an unstructured list of soil types (*e.g.* histosol, stagnosol) that is not related to the classes usually used for habitats (*e.g.* peat for histosol and wetland for stagnosol) and their properties (*e.g.* acidity or moisture).

**OntoBiotope ontology design**

We have built the OntoBiotope ontology according to the above criteria using a classical approach, assisted by automatic tools and ontology editors. Special attention is paid to terminology and formalism in order to allow its use for text indexing.

The OntoBiotope ontology Habitat section was built by following top-down and bottom strategies. The top-down approach divides the classification according to broad areas of study and their subdivisions. The bottom-up approach starts with all the particular terms that denote habitats and groups them iteratively and hierarchically. These terms were automatically extracted from the GOLD Habitat and GenBank source fields by the term extractor BioYaTeA (Golik et al., 2013). The manual terminology analysis of extracted terms was assisted by the TyDI (*Terminology Design Interface*) tool following the method described in (Nédellec et al., 2010). It consists in deciding for each term whether it represents a concept or is spurious or too precise, whether it is the best term to name a concept, or whether another form is preferable, whether the term is a term close to a term already used, in which case it is added as a synonym, or whether it represents a new habitat. The choice of the name of the term meets two criteria, it must be unambiguous (*e.g.* avoiding general word such as "system") and its use must be evidenced from domain corpora. These steps, which focus on adding new terms, alternate with tree reorganization steps.

---

[6] https://bioportal.bioontology.org/ontologies/ENVO

Sub-tree formalization is critical because it ensures that the properties of a class are shared by all its subclasses. By this way text information indexed by a particular class can be retrieved by querying with a more general class. Figure 1 shows, for example, the lineage of the yogurt class.

> yogurt / fermented dairy product / fermented food / processed food / prepared food / food for human / food.

FIGURE 1. Lineage of the yogurt class in OntoBiotope Habitat.

The first version of OntoBiotope Habitat distributed in 2010 has been extended according to the availability of new classifications and expertise opportunities. In particular, the Food sub-tree has been enriched using FoodEx, the new EFSA food classification and the expertise on positive food flora of the microbiologists of the Florilège project (Falentin et al., 2017).

**OntoBiotope Description**

The public version of OntoBiotope Habitat is distributed on the ontology portal AgroPortal[7] without any other relationship than the hierarchical relationship. It is available online and can be downloaded in standard OBO (original format) and RDF/XML formats. It contains 2,320 classes and 492 synonyms, organized in a hierarchy of a maximum depth of 13. The classes are supplemented by terminological information which is necessary for the automatic categorization of texts. Three types of synonyms are considered, exact, narrow and related. Exact synonyms are rare, they are mainly acronyms (e.g. *perchloroethylene contaminated site / PCE contaminated site*) or typographical variations. Narrow synonyms are used to avoid multiplying classes for irrelevant variations (e.g. *polluted site / contaminated site*). Related synonyms are used to refer to habitats that may eventually become classes in a later version (*PCP percolated soil/PCP contaminated soil*).

Ontobiotope Habitat root has 11 major domains (Figure 2). The distinction between so-called "artificial" and "natural" environments is subject to debate. The term "artificial" refers here to environments created and exploited by mankind (industrial, waste, housing). The term "natural" refers to environments that precede human intervention. "Living organism" and "Part of living organism" classes describes hosts and part of host the microorganisms interact with. The "Prepared food" branch is particularly developed because it has so far been little formalized in a microbiological context, with the notable exception of FoodEX2 classification, which inspired this work. OntoBiotope Habitat does not describe geographic locations because other classifications such as GeoNames base are dedicated to them.

**Examples of applications**

Ontobiotope Habitat has been used as a reference ontology for classifying microbial habitats in the three last editions of the international text-mining competition BioNLP Shared Task (2011, 2013 and 2016). In addition to stimulating text-mining research, OntoBiotope Habitat is used to automatically analyze and categorize microorganism habitats on a large scale by the Alvis text mining Suite that is deployed on the European OpenMinTeD text-mining infrastructure (Ba & Bossy, 2016). It equally deals with abstracts, full texts and database fields. The quality of Alvis Suite prediction results was measured through BioNLP Shared Task 2013 dataset where it achieved the best scores (Ratkovic et al., 2013).

Figure 3 shows an example of text-mining result (DOI: 10.1099/ijs. 0.63464-0). The taxon *Psychrobacter aquimaris* and habitat *sea water* are identified in the text, categorized and linked by the relationship *lives_in*.

---

[7] http://agroportal.lirmm.fr/ontologies/ONTOBIOTOPE

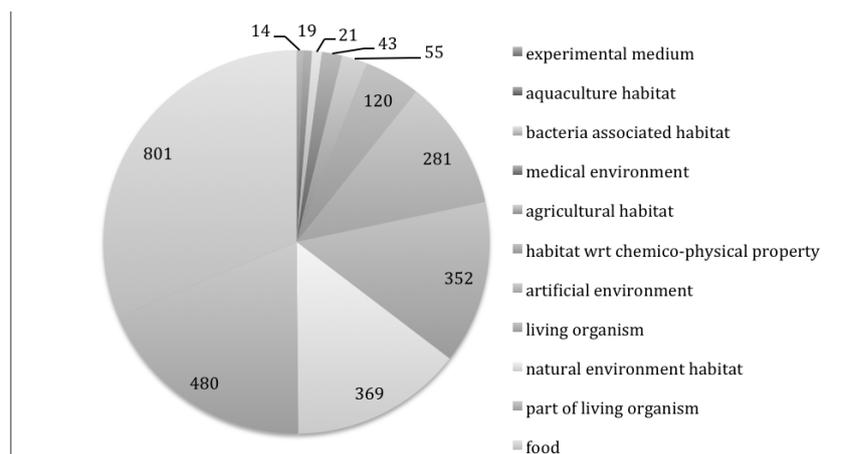

FIGURE 2. Distribution of OntoBiotope Habitats among upper classes.

Alvis used the NCBI taxinomy for taxon categorization. It assigned *marine water* category to *sea water* term that is itself a *marine environment* that is an *aquatic environment*, *etc.* according to OntoBiotope.

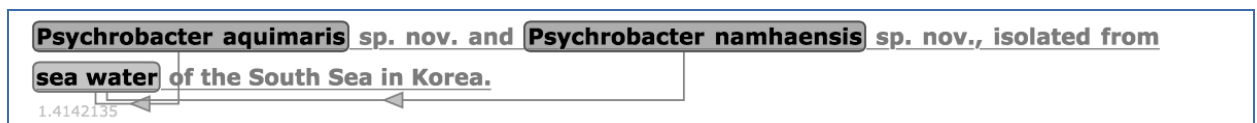

FIGURE 3. Screenshot of an example of textual information indexing by Alvis tool with OntoBiotope Habitat.

The categorization makes the text data searchable in an efficient way and comparable so that it can be extensively analyzed with other data sources. As an example, Table 1 shows the distribution of ubiquitous bacterium *Listeria monocytogenes* habitats in PubMed database. Alvis results are publicly accessible through the AlvisIR semantic search engine[8].

| Habitat | #freq | Habitat | #freq | Habitat | #freq |
|---|---|---|---|---|---|
| food | 1008 | dairy industry | 129 | patient | 74 |
| mouse | 797 | beef | 115 | fish | 71 |
| human | 511 | feces | 102 | murine | 69 |
| meat and meat product | 372 | milk and milk product | 96 | salmon | 65 |
| milk | 214 | salmon | 94 | breast | 64 |
| cheese | 181 | liver | 93 | turkey | 64 |
| ham | 153 | farm | 93 | gastrointestinal tract | 61 |
| animal | 149 | turkey | 86 | anaerobic environment | 59 |
| drinking water | 145 | liver | 81 | plant | 59 |
| spleen | 134 | fish | 77 | sausage | 58 |
| living organism | 133 | seafood and seafood product | 76 | patient | 56 |
| food processing factory | 132 | pork | 75 | raw milk | 55 |

TABLE 1. Top frequencies of OntoBiotope Habitat categories of *Listeria monocytogenes* in PubMed database.

**Perspectives**

OntoBiotope branches are at different stages of development, depending on the availability of existing ontologies that meet the need and collaboration with communities of experts. The living being and anatomy part is thus one of the parts that is richest in number of concepts (*e.g.* 51 classes in the gastrointestinal part class that all have been discovered in microbiology science documents), but its organization requires a more thorough investigation. Medical anatomy classifications may be not approprate since they are structured along functions but not according to the physico-chemical properties relevant to microorganisms such as oxygen availability. For example, it is more useful to relate urine to kidney or bladder from a microbial diversity perspective, rather than to body fluid, blood, saliva as the Medical Subject Headings thesaurus (MesH) does.

---
[8] http://bibliome.jouy.inra.fr/demo/ontobiotope/alvisir2/webapi/search

Several sub-trees, including habitat properties, microbial phenotypes and their industrial use are currently added to OntoBiotope to meet complementary needs that will soon be made publically available (Chaix et al., 2017).

**Keywords**
Ontology, text mining, microorganism, habitats.

**Acknowledgments**

This work was supported by the OpenMinTeD project (EC/H2020-EINFRA 654021). We thank the biologists of the Inra Florilège group for their contribution to the OntoBiotope ontology.

**References**
Ba M. and Bossy R. (2016). Interoperability of corpus processing workflow engines: the case of. AlvisNLP/ML in OpenMinTeD. *LREC Workshop on Cross-Platform Text Mining and Natural Language*, Portoroz.
Buttigieg P. L., Morrison N., Smith B., Mungall C. J., & Lewis S. E. (2013). The environment ontology: contextualising biological and biomedical entities.Journal of biomedical semantics, 4(1), 43.
Bossy, R., Jourde, J., Manine, A. P., Veber, P., Alphonse, E., Van De Guchte, M., ... & Nédellec, C. (2012). Bionlp Shared Task - the bacteria track. *BMC bioinformatics*, 13(11), S3.
Ivanova N., Tringe S. G, Liolios K., Liu W.-T., Morrison N., Hugenholtz P., and Kyrpides N. C. (2010). A call for standardized classification of metagenome projects. *Environmental microbiology*, 12(7):1803–1805.
Chaix, E., Aubin, S., Deléger, L., Nédellec, C. (2017). Text-mining needs of the food microbiology research community. *EFITA WCCA Congress*, Montpellier, FRA (2017-07-02 - 2017-07-06). http://prodinra.inra.fr/record/405408
Cook, H. V., Pafilis, E., & Jensen, L. J. (2016). A dictionary-and rule-based system for identification of bacteria and habitats in text. *ACL* 2016, 50.
Deléger, L., Bossy, R., Chaix, E., Ba, M., Ferré, A., Bessières, P., & Nédellec, C. (2016). Overview of the bacteria biotope task at BioNLP Shared Task 2016. In *Proceedings of the 4th BioNLP Shared Task Workshop* (pp. 12-22).
Falentin H., Chaix E., Derozier S., Weber M., Buchin S., Dridi B., ... Sicard D., Nédellec C. (2017) Florilège: a database gathering microbial phenotypes of food interest. *Microbial Diversity Conference*. Bari.
Cook, H. V., Pafilis, E., & Jensen, L. J. (2016). A dictionary-and rule-based system for identification of bacteria and habitats in text. BioNLP workshop, ACL 2016, 50.
Nédellec C., Nazarenko A. et Bossy R. (2009). "Information Extraction", *Handbook on Ontology*., S. Staab, R. Studer (eds.), Springer Verlag, p. 662-687.
Golik W., Bossy R., Ratkovic Z., Nédellec C. (2013) " Improving term extraction with linguistic analysis in the biomedical domain". Special Issue of the journal *Research in Computing Science,* vol 70 ISSN 1870-4069.
Nédellec, C., Golik, W., Aubin, S., & Bossy, R. (2010). Building large lexicalized ontologies from text: a use case in automatic indexing of biotechnology patents. *Knowledge Engineering and Management by the Masses*, 514-523.
Reddy T.B.K., Thomas A., Stamatis D., Bertsch J., Isbandi M., Jansson J., Mallajosyula J., Pagani I., Lobos E., Kyrpides N.C. (2014). The Genomes OnLine Database (GOLD) v.5: a metadata management system based on a four level (meta)genome project classification. *Nucleic Acids Res*.s;43:D1099–D1106.